\documentstyle[12pt]{article}

\begin{document}

\pagestyle{empty}

\begin{center}
{\bf MEASUREMENT AS ABSORPTION OF FEYNMAN TRAJECTORIES: COLLAPSE OF THE WAVE
FUNCTION CAN BE AVOIDED}

\medskip\medskip

by

\medskip

A. Marchewka

School of Physics and Astronomy

Tel-Aviv University, Tel-Aviv 69978, Israel

e-mail: marhavka@ccsg.tau.ac.il\\[3mm]

Z. Schuss

Department of Mathematics

Tel-Aviv University, Tel-Aviv 69978, Israel

e-mail: schuss@math.tau.ac.il\\[1cm]

{\bf ABSTRACT}
\end{center}

\vspace{3mm}
\noindent
We define a measuring device (detector) of the coordinate of quantum particle as
an absorbing wall that cuts off the particle's wave function. The wave
function in the presence of such detector vanishes on the detector.
The trace the absorbed particles leave on the detector is identifies
as the absorption current density on the detector. This density is
calculated from the solution of Schr\"odinger's equation with a
reflecting boundary at the detector. This current density is not the
usual Schr\"odinger current density. We define the probability
distribution of the time of arrival to a detector in terms of the
absorption current density. We define coordinate measurement by an
absorbing wall in terms of 4 postulates. We postulate, among others,
that a quantum particle has a trajectory. In the resulting theory the
quantum mechanical collapse of the wave function is replaced with the
usual collapse of the probability distribution after observation.
Two examples are presented, that of the slit experiment and the
slit experiment with absorbing boundaries to measure time of arrival.
A calculation is given of the two dimensional probability density
function of a free particle
from the measurement of the absorption current on two planes.

\newpage 
\pagestyle{plain}

\noindent
{\large {\bf 1. Introduction}}\newline
\renewcommand{\theequation}{1.\arabic{equation}}
\setcounter{equation}{0}
 
The concept of measurement is essential for the consistent mathematical
formulation of quantum mechanics. Several different theoretical approaches
to measurement exist in the quantum mechanics literature \cite{Bell}-\cite
{Gottfried}. 
The postulates of measurement theory in QM \cite{CT} describe a coordinate
measuring device as a machine that ``measures'' the coordinate by
collapsing instantaneously the wave
function to a point at a given moment of time. Repeated measurements of
identical particles result in a histogram of the coordinate at the given
time. According to QM this histogram is $\left| \Psi (x,t)\right| ^2$, where 
$\Psi (x,t)$ is the solution of Schr\"{o}dinger's equation.  
It is, however, universally agreed that
there is no known Hamiltonian that brings about such a sharp and
instantaneous transformation so that the description of this experiment may
not be contained in quantum mechanics. This quantum mechanical
collapse of the wave function is not the classical collapse of the
probability distribution of a random variable after it is observed.
The difference lies in the assumption that the coordinate of the
particle was undefined before the collapse of the wave function
occurred while the random variable had a definite value regardless of
whether it was observed or not.

This is the result of confusing the concept of the
coordinate of a particle at a given time with that of the probability
amplitude to find it there and then. Actually, there is no agreed
mathematical definition of what the instantaneous spatial coordinate of a
quantum particle is, although the expectation value is well defined.
Obviously, it is impossible to assign probability to an undefined event.

This confusion is illustrated by the following one-dimensional example.
Consider a particle on a linear segment bounded by two detectors. At the
initial time the wave function of the particle is prepared to be
symmetrically distributed about the center of the interval in the shape of
two packets traveling in opposite directions with equal momenta. According
to standard quantum mechanics (QM), at all times prior to detection the
wave function evolves according to Schr\"{o}dinger's equation, ignoring the
detectors, and assigns equal probabilities to detection at either end. The
two nearly separate packets are often interpreted, in the absence of a
better interpretation, as the particle being spread in various parts of the
interval up to the instance of detection. This interpretation is forced by
the QM view that a quantum particle does not have a trajectory 
\cite{Landau}. However, it is detected at only one end of the interval, 
resulting in the collapse of its wave function from a double packet 
to a delta function at one point. Could it be said that the other half 
of the packet reached that end in zero time?
This is one of the paradoxes that ensue from
the concept of collapse: the description of a particle as a wave
packet \cite{CT} implies that parts of the packet may move to the
detector in zero time. 

The interpretation of the square of the modulus of the wave function, 
$\left|\Psi (x,t)\right|^2$, as the histogram obtained from repeatedly
detecting the coordinate of identical particles, emitted under identical
conditions, at a given moment of time, $t$, ignores the detector and its
dynamics as part of the evolution of the particle. Ignoring the detector
makes the collapse of the wave function at the moment of detection
incompatible with the quantum description of the particle. Thus, the concept
of making a measurement without a measuring device is an oxymoron. Actually,
detecting particles by an impermeable screen, such as a photographic plate
or a fluorescent screen, cannot produce $\left| \Psi(x,t)\right|^2$ for a
particle in the entire space, because $\left|\Psi (x,t)\right|^2$
represents, according to quantum theory, the probability density of all
particles at $(x,t)$, including those that went across the screen and
returned to it at time $t$. Thus, in QM there is no probability density of
the points where particles appear on the detector, e.g., on a photographic
plate or a fluorescent screen. Thus, in one dimension,
the collapse machine is ``transparent to particles'' in the sense that  even if
the source of particles is on one side of the machine, it detects particles
whose wave function does not vanish on the other side of the machine.
In terms of Feynman trajectories, this means that Feynman trajectories 
can traverse the collapse machine in either direction. 

The concept of a quantum measuring device, as proposed by von
Neumann \cite{vonNeumann}, for example, does not resolve the above mentioned
problem. It is necessary to use a measuring device to observe the measuring
device thus putting the first quantum measuring device in contact with a
classical system that observes it. Therefore, observing one quantum system
with another does not resolve the problem of measurement. The quantum
measuring device, when observed, is in one and only one state while its wave
function spreads it among different states, as is the case with detecting a
particle \cite[Ch.1]{Grigolini}.

Furthermore, there is no definition of the notion of the time of arrival of
a particle at a detector (or any other point) \cite{vonNeumann}, \cite
{Salecker}-\cite{Aharonov}. These are all manifestations of the collapse
problem.

The same ambiguity arises with the concept of a probability current density 
\begin{equation}
{\cal J}(x,t)=2\Im m\bar{\Psi}(x,t)\nabla \Psi (x,t).  \label{netcurrent}
\end{equation}
It represents the net flow of particles at the point $x$ at time $t$ from
all directions. This is inconsistent with the concept of an impermeable or
absorbing screen (or detector) such as a photographic plate. The equation of
continuity implies, as in the case of diffusion \cite[Sect.5.4.2]{Gardiner}, 
\cite{Direct}, that the current on an absorbing screen must be unidirectional
and therefore cannot be given by eq.(\ref{netcurrent}). Thus a
unidirectional current has to be defined for the purpose of describing such
a measurement. Furthermore, on the one hand the observed pattern on an
absorbing screen is expected to show the unidirectional current density on
the screen, but on the other hand the postulates of measurement \cite
{Bell,Gottfried,CT} assert that what is detected is $\left| \Psi
(x,t)\right| ^2$. These two views seem to be incompatible with each other.

We consider now an ideal photographic plate or a fluorescent screen as a
measuring device (detector) of the coordinate of a quantum particle. 
The ideal plate is not transparent to particles
as can be seen from the experiment of putting two parallel plates one behind
the other. The plate on the side opposite to the source will  detect a
negligible fraction of the emitted particles. This is in contrast to the QM
collapse machine that is transparent in the sense that two collapse machines
in a row, when moved along the $x$-axis will produce the same histogram of a
packet of free particles, that is, their histograms are independent of one
another. 
A photographic plate differs from the QM collapse machine in other aspects
as well. First, it cuts the wave function by making it vanish on the
opposite side to that of the source, that is, it does not permit particles
through. Second, within the resolution of the observer, the plate records
continuously the time of arrival of particles. Although, upon the arrival of
a particle to the plate it instantaneously collapses the wave function to a
point mass, it is not a collapse machine in the sense of QM. The collapse
brought about by the detection of a particle in a plate is the same as that
in probability theory. That is, the probability distribution, that is spread
in space prior to detection, collapses into a point mass upon measurement. 

It can be assumed that an ideal photographic plate (or a grounded screen)
instantaneously absorbs the detected particles. Thus the measurement process
in such a device consists in irreversibly absorbing particles at a surface.
In addition, the $(t,x)$ histogram of arrivals produced by
moving the plate along the $x$-axis does not represent $\left| \Psi
(x,t)\right|^2$ of the freely propagating packet, but rather the 
{\em unidirectional absorption flux density} on the plate.

It is the purpose of this paper to replace the postulates of QM by a set of
postulates that contain an absorbing wall (detector). To this end, we adopt
Feynman's formulation of QM in terms of trajectories and the Feynman
integral. The postulates assert that\\

\noindent{\em 1. A quantum particle has a trajectory}\\

\noindent{\em 2.} {\em A detector of spatial coordinate is a surface 
that absorbs Feynman trajectories when they reach it for the first time}.\\

\noindent{\em 3. At any time the wave function of a quantum particle in the
presence of an absorbing wall is the Feynman integral over all trajectories
that have not reached the absorbing wall by that time.}\\

\noindent{\em 4. At the time of arrival the wave function of the particle is
described by the absorbing surface (not by Feynman's integral).}\\

Postulate 1 implies that the trajectories of a quantum particle are Feynman
trajectories.
The view expressed in \cite{Landau}, that an electron cannot have a
trajectory, is based on the argument that tighter measurements of the
coordinate of an electron at fixed time intervals lead to wilder
vacillations in the position of the electron. This view requires 
revision. According
to \cite{Landau}, the measurement process interacts with the electron, thus
modifying its trajectory. The tighter is the measurement the stronger is the
interaction. Thus tighter measurements measure different trajectories than
do looser measurements. Consider for example a measurement by a circular
hole in a wall followed by a photographic plate. The smaller is the hole the
more spread is the distribution of the spots that electrons filtered through
the hole leave on the plate. This, contrary to the view expressed in 
\cite[Ch.1,\S7]{Landau}, does not imply that an electron has no trajectory,
because the measurements by various size holes measure different
trajectories, due to the different interactions of the different holes with
the electron. This is an expression of the uncertainty principle. It cannot
be argued that the same trajectory passes, unmodified, through the various
size holes, because, as stated in \cite[Ch.1,\S1]{Landau}, the influence
of the smaller hole on the electron (and therefore on its trajectory, if it
has one) is stronger than that of the bigger hole.

In \cite[Ch.1, \S 1]{Landau} it is stated that trajectory (coordinates)
measurements at shorter time intervals apart gives closer results for the
coordinate measurement. This is compatible with the notion of continuous
trajectories of the electron.

Postulate 2 does not imply that the absorbed particle disappears. It should be
understood in the following sense. Once a particle has reached the absorbing
wall its (Feynman) trajectory is terminated, but a new trajectory begins,
possibly at the same place and the future evolution of the wave function is
subject to another Hamiltonian. The initial value of the wave function after
absorption may be concentrated at the terminal position of the particle on
the absorbing wall. As described in \cite{PLA}-\cite{PhD}, Feynman
trajectories that have been absorbed in the absorbing boundary not
longer interfere with those that have not been absorbed so far. This
means that the Feynman integrals over the surviving trajectories and
over the absorbed trajectories are orthogonal. That is, setting
\begin{eqnarray*}
\psi_{\mbox{surviving}}(x,t)=
\int_{\stackrel{\mbox{surviving}} {\mbox{trajectories}}}
\exp\left\{\frac{i}{\hbar}S[x(\cdot),t)]\right\}\,{\cal D}x(\cdot)\\
\psi_{\mbox{absorbed}}(x,t)=\int_{\stackrel{\mbox{absorbed}}
{\mbox{trajectories}}}
\exp\left\{\frac{i}{\hbar}S[x(\cdot),t)]\right\}\,{\cal D}x(\cdot),
\end{eqnarray*}
we assume that for all $t$
\begin{eqnarray}
\int \Re{\mbox{e}}\left\{\psi_{\mbox{surviving}}(x,t)
\bar{\psi}_{\mbox{absorbed}}(x,t)\right\}\,dx=0.\label{separation}
\end{eqnarray}
We call eq.(\ref{separation}) {\em a separation principle}.

According to Postulate 3, $\left| \psi(x,t)\right| ^2$ represents the 
{\em conditional probability density} that the particle is at position $x$ at 
time $t$, given that it has not reached the absorbing wall 
by time $t$. It was shown in \cite{PLA} and \cite{PAPER2} 
that postulates 2 and 3 imply that $\left|\psi (x,t)\right|^2$ is the
solution of Schr\"odinger's equation with zero boundary conditions on
the absorbing wall. The {\em joint probability density} that the
particle is at position $x$ at time $t$ and the particle has not
reached the absorbing boundary by time $t$ is 
$|\Psi(x,t)|^2=S(t)|\psi(x,t)|^2$, where $S(t)$ is the {\em survival
probability} of the particle. That is, $S(t)$ is the probability that
the particle has not reached the absorbing boundary by time $t$. The
survival probability $S(t)$ has been calculated in \cite{PLA} and
\cite{PAPER2} and is discussed below in the context of the concept of
{\em time of arrival}. The histogram constructed on the 
detector (the absorbing wall) is neither $\left|\psi (x,t)\right|^2$
nor $|\Psi(x,t)|^2$, but rather the unidirectional current
density of the absorbed Feynman trajectories (particles), that is, the
absorption current density.

While $\psi(x,t)$ is obtained from a Hamiltonian theory with infinite
potential behind the wall, the discounted wave function
$|\Psi(x,t)|^2=S(t)|\psi(x,t)|^2$ is not. It does not preserve
probability in the domain bounded by the absorbing boundary because
\[\int |\Psi(x,t)|^2\,dx=S(t)\int|\psi(x,t)|^2\,dx=S(t),\]
which decays in time. Obviously, if the absorbed Feynman trajectories
are tracked after absorption (i.e., are included in the Feynman
integral, but without interference with the trajectories that have not
been absorbed so far), total probability is preserved. This is in
contrast to complex valued Hamiltonians that lead to loss of total
probability.

According to postulate 4, the joint statistics of the location of the point
where the particle arrives at the detector and the time of arrival are
determined, as shown below, by the unidirectional probability current
density on the wall, not by $\left| \psi (x,t)\right| ^2$ at the wall. Thus,
in one dimension, the probability density that the particle arrives at the
detector placed at $x=0$ at time $t$ is given by eqs.(\ref{Jlambda}) and (%
\ref{psi}), whereas QM predicts that the density is $\left| \Psi
(0,t)\right| ^2$. In higher dimensions the statistics of the point and time
of arrival of the particle on the detector (wall) are given by eqs.(\ref
{Jblambda}) and (\ref{PSI}), whereas QM predicts $\left| \Psi ({\bf x}%
,t)\right| ^2$ on the detector. If the surface of a two dimensional detector
is the plane $x=0$, the quotient of the two statistics is only a function of 
$t$. That is, both theories predict the same pattern on the detector screen
at any time $t$. According to QM, there is no definition of the pattern
obtained by repeatedly firing electrons at the screen because there is no
definition of the time of arrival. Instead, this pattern is approximated by $%
\left| \Psi ({\bf x},t)\right| ^2$ on the detector at some mean time
obtained from semi-classical considerations \cite{Feynman}. According to the
theory obtained from the above postulates, the pattern is obtained by
integrating the flux density on the screen over all times.

We refer to the quantum mechanics defined by the above postulates 
{\em measurable quantum mechanics} (MQM). It is obvious, that QM is
recovered from MQM by moving the absorbing wall to infinity. This renders
MQM and extension of QM. In MQM the Feynman trajectories are interpreted as
the possible trajectories of a particle. The description of a particle in
MQM adopts the language of stochastic processes: a particle is the set of
all continuous functions (trajectories), a $\sigma$-algebra of measurable
sets of trajectories, and a Feynman integral defined on measurable
sets of trajectories. This is analogous to the
theory of diffusion processes and their sample paths \cite{Ito} which
defines a diffusing particle (in the sense of a stochastic process) as the
set of all continuous functions (trajectories), a $\sigma$-algebra of
measurable sets of trajectories, and a Wiener integral. It so happens that
the Wiener integral defines a measure in function space and can be
applied in particular to diffusion theory with absorbing boundaries 
\cite{Gardiner,Direct,book}. The Feynman integral is countably
additive set function and is therefore a vector valued measure. The
probability defined by the Feynman integral, however, is not a
countably additive set function. This is the feature of the Feynman
integral that gives rise to interference. Thus, interference pattern
is observed in the double slit experiment when the particles are sent
one at a time.

The problem of absorption in quantum systems was considered by us in \cite
{PLA} and \cite{PAPER2}. Absorbing boundaries in a finite interval
$[a,b]$ are described by the following assumptions:\\
\noindent
{1. \em A trajectory that reaches $x=a$ or $x=b$
is instantaneously absorbed. \\
2. The population inside the interval $(a,b)$ is reduced
by the probability absorbed at the boundary. \\
3.  The absorption process at time $t$ is the limiting
process as $\Delta t\rightarrow 0$ of absorbing trajectories that
survived in the interval $\left[ a,b\right] $ till time $t$, and
propagated
into the boundary in the time interval $\left[ t,t+\Delta t\right]
$.\\
4. The probability of the absorbed trajectories in the time interval
$[t,t+\Delta t]$ is proportional to the p.d.f. to
propagate into the boundary in this time interval. The proportionality
constant is a characteristic length.}

It was shown in \cite{PLA} that for an absorbing wall at the origin, the
survival probability of a quantum particle with an absorbing wall (detector)
at the origin is given by 
\begin{equation}
S(t)=\exp\left\{-\frac{\lambda\hbar}{m\pi}\int_0^t\,\left|\,
\frac{\partial\psi_B(0,t')}{\partial x}\right|^2\,dt'\right\},\label{psi}
\end{equation}
where $\psi_B(x,t)$ is the wave function obtained from Schr\"{o}dinger's
equation with a reflecting wall at the origin (zero boundary conditions at 
$x=0$ or an infinite potential behind the wall). Actually, $\psi _B(x,t)=0$
for $x\geq 0$ and $\psi _B(x,t)\neq 0$ for $x<0$.\\

\noindent
{\large{\bf 2. A mathematical model of quantum mechanics with measurements}}\\
\renewcommand{\theequation}{2.\arabic{equation}} \setcounter{equation}{0}

In Bayesian probability theory \cite{DeFinetti} it is impossible to assign
probability to events that cannot be observed. The assignment of
probabilities to events is based on prior observations. Otherwise, all
events are equally likely. Thus measurements (observations) are the basis
for constructing a probability theory that describes any experiment. These
facts apply to quantum mechanics (QM) as well. The most basic concepts of QM
are probability amplitude and probability density. These have never been
measured under the condition of absence of a measuring device (e.g.,
photographic plate, fluorescent screen, etc.) whereas the measuring device
always influences the electron in a drastic way. Furthermore, according to
standard quantum mechanics (QM), the measurement process influences the
wave function in a drastic way as well, namely, the wave function
``collapses''.

Thus, to be mathematically consistent, the accumulated data from measuring
the location of a single electron as a point in time and space cannot be
used for the construction of QM in which a measuring device is completely
neglected, as is common practise in QM. Therefore, the introduction of a
mathematical model of a measuring device in the mathematical formulation of
QM is inescapable.

The fundamental requirement of a mathematical model of a quantum measuring
device (QMD) is that it reflects what is actually measured. Thus, we define
a QMD (for the purpose of this discussion) as a device that measures the
electron as a point in time and space. Our postulate is that prior to
measurement the electron is described by the wave function and at the moment
of measurement it is described by the QMD. This reflects our uncertainty
about the location of the electron as long as it has not been observed,
however, when it is observed, our uncertainty is instantaneously replaced with
certainty. Note that this statement implies that the electron actually has a
location at all times prior to measurement, which in turn implies that it
actually has a trajectory, in contravention to the commonly held view in QM 
\cite{Landau}. This difference, between the description of a random variable
or random process prior to observation and at observation time, is
fundamental in probability theory and therefore should be no surprise in QM.
Obviously, this formulation of the mathematical model of quantum mechanics
eliminates the notion of collapse of the wave function.

In view of the above discussion, we adopt Feynman's formulation of QM,
which is based on the concept of trajectories, although the Feynman
trajectories have not been interpreted so far as actual trajectories of
electrons. Feynman's formulation of QM is equivalent to Schr\"{o}dinger's
formulation and therefore suffers from the same difficulties concerning the
concept of measurement. Our interpretation of Feynman's QM is as follows.
The mathematical model of an electron is a stochastic (random) process with
continuous trajectories with a vector valued density, called the wave
function. According to the standard methodology of probability theory \cite
{Doob}, such a random process consists of the space of trajectories (all
continuous functions of time with values in ${\bf R}^3$)\newline

\noindent
{\large{\bf 3. Trajectories of a diffusion process}}\newline
\renewcommand{\theequation}{3.\arabic{equation}} \setcounter{equation}{0}

We begin with a brief review of relevant notions from diffusion theory. In
diffusion theory \cite{Ito} the motion of a Brownian particle in a domain $%
{\cal D}\,$is modeled as a stochastic process. That is, by definition, a
Brownian particle is, mathematically, the set of all trajectories, $\Omega $%
, (all continuous paths) and a Wiener integral defined on them (or a Wiener
measure defined on the sample space of continuous functions). It can be
shown that the Wiener integral assigns probability zero to all trajectories
that have a finite velocity at any time. The Wiener integral over all
trajectories, $\omega =$ ${\bf x}(t)$, that start at time zero at the point $%
{\bf x}_0$ and reach the point ${\bf x} $ at time $t$, denoted $p({\bf x}%
,t\,|{\bf \,x}_0)$, satisfies the Fokker-Planck equation 
\[
\frac \partial {\partial t}p({\bf x},t\,|\,{\bf x}_0)=-\nabla \cdot {\cal J}(%
{\bf x},t\,|\,{\bf x}_0), 
\]
where ${\cal J}$ is the probability current density of trajectories at the
point ${\bf x}$ at time $t$, that started out at{\bf \ }${\bf x}_0$. At an
absorbing boundary,$\ \Gamma $, say, $\left. p({\bf x},t\,|\,{\bf x}%
_0)\right| _\Gamma =0$, and $\left. {\cal J}({\bf x},t\,|\,{\bf x}_0)\right|
_\Gamma $ is the uni-directional current of trajectories into the absorbing
boundary $\Gamma $ \cite{Direct}. The function 
\begin{equation}
p({\bf x}\,|\,{\bf x}_0)=\int_0^\infty p({\bf x},t\,|\,{\bf x}_0)\,dt
\label{pxx0}
\end{equation}
is the mean time trajectories that start out at ${\bf x}_0$ spend at ${\bf x}
$ prior to absorption. It satisfies the stationary Fokker-Planck equation
with a source at ${\bf x}_0$ and vanishes on $\Gamma $.

Denoting by $\tau (\omega )$ the first passage time of a trajectory $\omega
\in \Omega $ to $\Gamma $ (thus $\tau (\omega )$ is a random variable
defined on trajectories), we have 
\begin{equation}
\Pr \left\{ \tau (\omega )>t\,|\,{\bf x}_0\right\} =\int_{{\cal D}}p({\bf x}%
,t\,|\,{\bf x}_0)\,\,d{\bf x.}  \label{Ptaut}
\end{equation}
It follows from the Fokker-Planck equation that 
\begin{equation}
\Pr \left\{ \tau (\omega )=t\,|\,{\bf x}_0\right\} =\oint_\Gamma {\cal J}(%
{\bf x},t\,|\,{\bf x}_0)\cdot {\bf n(x})\,dS_{{\bf x}},  \label{taudensity}
\end{equation}
where ${\bf n(x})$ is the unit outer normal to $\Gamma $ at the point ${\bf x%
}$ and $dS_{{\bf x}}$ is a surface area element on $\Gamma $. Thus, 
\[
\oint_\Gamma {\cal J}({\bf x},t\,|\,{\bf x}_0)\cdot {\bf n(x})\,dS_{{\bf x}} 
\]
is the probability density function of the first passage time of a
trajectory to $\Gamma $. That is, the normal component of $\left. {\cal J}(%
{\bf x},t\,|\,{\bf x}_0)\right| _\Gamma $ is the probability density of
points where Brownian trajectories hit the boundary at time $t\,\,$(for the
first and last time!).

If a Brownian particle is released at the point ${\bf x}_0$ at time $t=0$
and is detected (observed) for the first time when it reaches that absorbing
boundary $\Gamma $, it appears there as a single point. The probability
density function of the points on $\Gamma $ where the particle is observed
(measured, absorbed) at time $t$ (per unit area and unit time) is 
\[
\Pr \left\{ {\bf x(}t)={\bf x,\,}\tau (\omega )=t\,|\,{\bf x}_0\right\} =%
{\cal J}({\bf x},t\,|\,{\bf x}_0)\cdot {\bf n(x}). 
\]
Thus the normal component of the probability flux density represents the
joint probability density of two events: the time of arrival of the measured
particle is $t$ and the point where it appears on the screen is ${\bf x}$.
The probability density function of the points where it leaves a trace on
the detector (a photographic plate, say) is 
\[
\Pr \left\{ {\bf x(}\tau (\omega ))={\bf x}\,|\,{\bf x}_0\right\} ={\cal J}(%
{\bf x}\,|\,{\bf x}_0)\cdot {\bf n(x)=}\int_0^\infty {\cal J}({\bf x},t\,|\,%
{\bf x}_0)\cdot {\bf n(x})\,dt, 
\]
where ${\cal J}({\bf x}\,|\,{\bf x}_0)$ is the current density calculated
from the function $p({\bf x}\,|\,{\bf x}_0)$ (see eq.(\ref{pxx0})). All
these densities can be constructed from histograms of the outcomes of
repeated identical experiments.

We adopt the same approach for a quantum particle, with the only difference
that the Wiener integral is replaced with the Feynman integral. All these
notions are generalized to Feynman integrals (see \cite{PLA}, 
\cite{PAPER2}, \cite{PAPER1}), though their calculation is different 
from that of their Wiener integral counterparts.\newline

\noindent
{\large{\bf 4. Uni-directional current and time of arrival}}\newline
\renewcommand{\theequation}{4.\arabic{equation}}\setcounter{equation}{0}

In order to complete the construction of the description of measurement, we
introduce the concept of a unidirectional probability current density. Since
the wave function vanishes at and beyond the absorbing wall \cite{PLA} the
Schr\"{o}dinger current vanishes there. We define, therefore, the
instantaneous unidirectional current into an absorbing wall as the square of
the modulus of the Feynman integral over trajectories that propagate into
the wall per unit time. This notion can be generalized to the situation
where the wave function or its gradient suffer a discontinuity across a wall 
\cite{PhD}. It was shown in \cite{PAPER1} that the uni-directional current
at an absorbing wall at the origin (in one dimension) is given by
\begin{equation}
{\cal J}(0,t)=\frac{\lambda\hbar}{m\pi}\left|\frac{\partial\psi(0,t)}
{\partial x}\right|^2=\frac{\lambda\hbar}{m\pi}\left|\frac{\partial
\psi_B(0,t)}{\partial x}\right|^2\exp\left\{-\frac{\lambda\hbar}{m\pi}
\int_0^t\left|\frac{\partial\psi_B(0,t')}{\partial x}\right|^2\,dt'
\right\}\label{Jlambda}
\end{equation}
(see eq.(\ref{psi})). It represents the probability per unit time $\Delta t$
of Feynman trajectories that propagate in the time interval $[t,t+\Delta t]$
into an point $(x=0)$ in the wall. In diffusion theory this definition gives
the usual expression for the probability current at an absorbing boundary 
\cite{Gardiner,Direct}.

It was shown in \cite{PLA,PAPER2} that the probability distribution of the time of
arrival of a particle at a detector is determined by the relation 
\[
\Pr \left\{ \tau >t\right\} =\int_0^\infty \ \left| \psi (x,t)\,\right|
^2dx\,\ =\exp \left\{ -\frac{\lambda\hbar}{m\pi} \int_0^t\left| \frac{\partial \psi
_B(0,t^{\prime })}{\partial x}\right| ^2\,\ dt^{\prime }\right\} . 
\]
In the experiment of measuring the time of arrival of a particle at a
detector, $\tau $, the information of non-arrival of the particle at the
detector is available continuously all the time. Thus, the discounting in
the time interval $[t,t+\Delta t]$ has to be conditioned on the information
that the particle has not arrived at the detector prior to time $t$, that
is, on the event $\left\{ \tau >t\right\} $. The conditional probability is
given by 
\[
\Pr \left\{ t<\tau <t+\Delta t\,|\,\tau >t\right\} =\frac{\Pr \left\{ t<\tau
<t+\Delta t\right\} }{\Pr \left\{ \,\tau \geq t\right\} }=\frac{\lambda\hbar}{m\pi} \Delta
t\left| \frac{\partial \psi _B(0,t)}{\partial x}\right| ^2+o\left( \Delta
t\right) . 
\]
At each time $s$ prior to the arrival of the particle at the detector the
probability distribution of the arrival time, $\tau $, is conditioned on $%
\tau \geq s$. That is, for $t\geq s$, 
\[
\Pr \left\{ \tau >t\,|\,\tau >s\right\} =\frac{\Pr \left\{ \tau \geq t\,\cap
\,\tau \geq s\right\} }{\Pr \left\{ \,\tau \geq s\right\} }=\frac{\Pr
\left\{ \,\tau \geq t\right\} }{\Pr \left\{ \,\tau \geq s\right\} }=\exp
\left\{ -\frac{\lambda\hbar}{m\pi} \int_s^t\left| \frac{\partial \psi _B(0,t^{\prime })}{%
\partial x}\right| ^2\,\ dt^{\prime }\right\} . 
\]
It follows that the conditional probability density of the time of arrival
of the watched detector for any time $t\geq s$ is 
\begin{eqnarray}
\Pr \left\{ \tau =t\,|\,\tau \geq s\right\} =\frac{\lambda\hbar}{m\pi} \left| \frac{\partial
\psi _B(0,t)}{\partial x}\right| ^2\exp \left\{ -\frac{\lambda\hbar}{m\pi} \int_s^t\left| 
\frac{\partial \psi _B(0,t^{\prime })}{\partial x}\right| ^2\,\ dt^{\prime
}\right\} . \label{PtauB}
\end{eqnarray}
Hence, the rate of arrival of the particle at the detector at time $s$ is 
\[
f_\tau (s)=\lim_{t\rightarrow s}\Pr \left\{ \tau =t\,|\,\tau \geq s\right\}
=\frac{\lambda\hbar}{m\pi} \left| \frac{\partial \psi _B(0,s)}{\partial x}\right| ^2. 
\]

As an application of this theory, we consider the following experiment. A
particle is started with initial wave function $\psi (x,0)$ and a detector
placed at the origin. The detector records the time between the release of
the particle and its arrival at the detector. Repeated recordings of
identical experiments form a histogram of the times of arrival. Adopting
the approach that the recordings and the histogram are independent of
whether the detector is watched or not, we calculate below the probability
distribution obtained from the histogram in the limit of large number of
experiments. According to the adopted approach, we assume that the detector
is watched continuously. Thus information of non-arrival of the particle at
the detector is available continuously. 

The measurement of the point of arrival of a quantum particle at a detector
requires a higher dimensional formulation. In higher dimensions the
discounted wave function in a domain ${\cal D}${\em \ } in the presence of
an absorbing boundary $\Gamma $ is given by 
\begin{equation}
\psi ({\bf x},t)=\psi _B({\bf x},t)\exp \left\{ -\frac{\lambda\hbar}{2
m\pi}\int_0^t\
\oint_\Gamma \left| \frac{\partial \psi _B({\bf x}^{\prime },t^{\prime })}{%
\partial {\bf n}}\right| ^2\,\,dS_{{\bf x}^{\prime }}\,dt^{\prime }\right\} ,
\label{PSI}
\end{equation}
where $\psi _B({\bf x},t)$ is the solution of Schr\"{o}dinger's
equation in ${\cal D}$ with zero boundary condition on $\Gamma $. Adopting
the interpretation of the squared modulus of the wave function as the
probability density of finding a particle at the point ${\bf x}$ at time $t$
(whatever that means), the discounted wave function can be used to calculate
the {\em joint} probability density of surviving by time $t\,$and finding
the particle at ${\bf x}$ at the same time. The squared modulus of the wave
function, {\em conditioned} on surviving by time $t$ is found by dividing
the joint probability density $\left| \psi ({\bf x},t)\right| ^2$ by the
probability of the condition, $S(t)$. In the multi-dimensional case at hand 
\begin{equation}
S(t)=\exp \left\{ -\frac{\lambda\hbar}{m\pi} \int_0^t\ \oint_\Gamma \left| \frac{\partial \psi
_B({\bf x}^{\prime },t^{\prime })}{\partial {\bf n}}\right| ^2\,\,dS_{{\bf x}%
^{\prime }}\,dt^{\prime }\right\} .  \label{mdSt}
\end{equation}
It follows from eqs.(\ref{PSI}) and (\ref{mdSt}) that the conditioned wave
function is $\psi _B({\bf x},t)$.

Applying the method of propagation into the wall and the discounting
procedure, as in \cite{PLA}, we find that the normal component of the
multi-dimensional probability current density at any point ${\bf x}$ on $%
\Gamma $ is given by 
\begin{equation}
\left. {\cal J}({\bf x},t)\cdot {\bf n}\left( {\bf x}\right) \right| _\Gamma
=\frac{\lambda\hbar}{m\pi} \left| \frac{\partial \psi _B({\bf x},t)}{\partial {\bf n}}\right|
_\Gamma ^2\,\exp \left\{ -\frac{\lambda\hbar}{m\pi} \int_0^t\oint_\Gamma \left| \frac{\partial
\psi _B({\bf x}^{\prime },t^{\prime })}{\partial {\bf n}}\right| ^2\,\,dS_{%
{\bf x}^{\prime }}\,\,dt^{\prime }\right\} .  \label{JnG}
\end{equation}
As in the case of diffusion, the pdf of the arrival time is given by 
\begin{equation}
\Pr \left\{ \tau =t\,\right\} =\oint_\Gamma {\cal J}({\bf x},t\,)\cdot {\bf %
n(x})\,dS_{{\bf x}}.  \label{PtautF}
\end{equation}
The probability density function of the particle hitting the point ${\bf x}$
on the detector is given by 
\begin{equation}
\Pr \left\{ {\bf x(}\tau )={\bf x}\right\} ={\cal J}({\bf x}\,)\cdot {\bf %
n(x)=}\int_0^\infty {\cal J}({\bf x},t\,)\cdot {\bf n(x})\,dt.
\label{PxtautF}
\end{equation}
Obviously, the integral converges and 
\[
\oint_\Gamma \Pr \left\{ {\bf x(}\tau )={\bf x}\right\} \,dS_{{\bf x}%
^{\prime }}=1-\exp \left\{ -\frac{\lambda\hbar}{m\pi} \int_0^\infty \oint_\Gamma \left| \frac{%
\partial \psi _B({\bf x}^{\prime },t^{\prime })}{\partial {\bf n}}\right|
^2\,\,dS_{{\bf x}^{\prime }}\,\,dt^{\prime }\right\} . 
\]
Thus, if 
\[
\int_0^\infty \oint_\Gamma \left| \frac{\partial \psi _B({\bf x}^{\prime
},t^{\prime })}{\partial {\bf n}}\right| ^2\,\,dS_{{\bf x}^{\prime
}}\,\,dt^{\prime }=\infty , 
\]
then 
\begin{equation}
\Pr \left\{ \tau <\infty \,\right\} =\oint_\Gamma \Pr \left\{ {\bf x(}\tau )=%
{\bf x}\right\} \,dS_{{\bf x}^{\prime }}=1,  \label{pTAUINF}
\end{equation}
which means that the particle is absorbed in finite time with probability 1.
However, if 
\[
\int_0^\infty \oint_\Gamma \left| \frac{\partial \psi _B({\bf x}^{\prime
},t^{\prime })}{\partial {\bf n}}\right| ^2\,\,dS_{{\bf x}^{\prime
}}\,\,dt^{\prime }<\infty , 
\]
then 
\begin{equation}
\Pr \left\{ \tau <\infty \right\} =\oint_\Gamma \Pr \left\{ {\bf x(}\tau )=%
{\bf x}\right\} \,dS_{{\bf x}^{\prime }}<1,  \label{pTAUFIN}
\end{equation}
so that 
\[
\Pr \left\{ \tau =\infty \right\} >0. 
\]
That is, there is finite probability that the particle is never absorbed at
the detector.\newline

\noindent
{\large{\bf 5. Collapse of the wave function and measurement}}\newline
\renewcommand{\theequation}{5.\arabic{equation}}\setcounter{equation}{0}

We begin the discussion of the collapse of the wave function with the
analysis of an analogous situation in diffusion theory, where this concept
has been well understood for a long time now. A Brownian particle released
at a point appears (is observed) as a point at a random time and at a random
coordinate on an absorbing wall that is used as a detector. Its probability
density function evolves according to the Fokker-Planck equation which takes
into account the absorbing wall as a boundary condition: it vanishes there
at all times \cite{Gardiner,Direct,book}. When the Brownian particle (its
trajectory) reaches the wall it is observed as a point on the wall so that its
probability density function collapses instantaneously to a delta function
there and then. There is no mechanism in the Fokker-Planck equation to
effect this sudden catastrophe and there cannot be one. This apparent
paradox is resolved through the connection between the Brownian trajectories
and the Wiener integral (the solution of the Fokker-Planck equation). This
probability density function represents our uncertainty about the location
of the Brownian trajectory at all times prior to absorption in the wall and
assigns densities to all possible Brownian trajectories. It assigns a
probability density to the point and time of arrival of the Brownian
particle (that is, its trajectory) at the wall by means of the probability
current density on the wall, as described in Section 2 above. This density
on the wall can be constructed from a histogram of the points of arrival of
identical Brownian particles, as mentioned there.

According to MQM, the situation with a quantum particle is quite similar. For a
one-dimensional quantum particle in the presence of detectors at the ends of
a given interval (absorbing walls, say), as long as its trajectory has not
reached either detector its wave function evolves according to the
Schr\"{o}dinger equation with boundary conditions given at the endpoints of
the interval, where it vanishes at all times \cite{PLA,PAPER1,PhD} and is
discounted as described in Section 1 and given by eq.(\ref{psi}). 
When the particle (its trajectory) reaches either end it is observed 
there. Prior
to this observation the particle's wave function assigns a probability
amplitude to all possible Feynman trajectories that have not reached the
boundaries. Actually, it vanishes outside the interval at all times. The
wave function assigns a probability density to the point and time of arrival
of the quantum particle (that is, of its trajectory) at the endpoints by
means of the probability current density there, as described in Sections 3,4
above.

In summary, a quantum particle (that is, its trajectory) arrives at a
detector with the probability density defined by the uni-directional current
at the detector, computed from the discounted wave function that is
constructed from the solution of the Schr\"{o}dinger equation with zero
boundary conditions at the detector by eq.(\ref{psi}). This description
eliminates the need for the notion of the QM collapse of the wave function.
The QM collapse is replaced in MQM by the usual collapse of
probability theory.

In higher dimensions the wave function in a domain ${\cal D}$
in the presence of an absorbing boundary $\Gamma $ is given by 
\begin{equation}
\psi({\bf x},t)=
\psi_B({\bf x},t)\exp\left\{-\frac{\lambda\hbar}{2m\pi}\int_0^t\left|
\frac{\partial\psi_B({\bf x},t')}{\partial {\bf n}}\right|
_\Gamma ^2\,dt^{\prime }\right\} ,  \label{PSIB}
\end{equation}
where $\psi _B({\bf x},t)$ is the solution of Schr\"{o}dinger's equation in $%
{\cal D}$ with zero boundary condition on $\Gamma $. The probability current
density on $\Gamma $ is given by 
\begin{eqnarray}
\left.{\cal J}({\bf x},t)\right|_\Gamma&=&\frac{\lambda\hbar}{m\pi}\left| 
\frac{\partial\psi({\bf x},t)}{\partial{\bf n}}\right|_\Gamma ^2\nonumber\\
&=&
\frac{\lambda\hbar}{m\pi}\left|\frac{\partial\psi_B({\bf x},t')}
{\partial {\bf n}}\right|_\Gamma^2\,\exp\left\{ -\frac{\lambda\hbar}{m\pi} \int_0^t\ \left| \frac{\partial \psi _B(%
{\bf x},t^{\prime })}{\partial {\bf n}}\right| _\Gamma ^2\,\,dt^{\prime
}\right\} .  \label{Jblambda}
\end{eqnarray}
The notions of current density at $\Gamma $ and time of arrival are defined
in an analogous manner, as described in Sections 2-4. This is illustrated in
the slit experiment application below.\\

\noindent
{\large{\bf 6. Application to the slit experiment}}\newline
\renewcommand{\theequation}{6.\arabic{equation}}\setcounter{equation}{0}

We consider the following experimental setup. A planar screen is placed in
the plane $x=0$ and another screen it placed in the plane $x=x_0$ and it is
slit along a line parallel to the $z-$axis. Due to the invariance of the
geometry of the problem in $z$ the mathematical description of the slit is,
following \cite{Feynman}, an initial truncated Gaussian wave packet in the $%
(x,y)-$plane, concentrated around the initial point, $(x_0,0).$ To describe
the interference pattern on the screen, we assume it is an absorbing line on
the $y$-axis and apply the formalism developed above. The wave function, as
given by our formalism, evolves from the initial packet according to eq.(\ref
{PSIB}), as 
\[
\psi (x,y,t)=\psi _B(x,y,t)\exp \left\{ -\frac{\lambda\hbar}{2m\pi}\int_0^t\
\int_{-\infty }^\infty \left| \frac{\partial \psi _B(0,y^{\prime },t^{\prime
})}{\partial x}\right| ^2\,\,dy^{\prime }\,dt^{\prime }\right\} \,, 
\]
where $\psi _B(x,y,t)$ is the solution of the Schr\"{o}dinger equation in
the half plane $x>0$ with $\psi _B(0,y,t)=0$ and $\psi _B(x,y,0)$ is the
given initial packet. The probability distribution of the time to arrival of
the Feynman trajectories to the absorbing line is determined from the
equation 
\[
\Pr \left\{ \tau >t\right\} =\int_0^\infty \int_{-\infty }^\infty \left|
\psi (x,y,t)\,\right| ^2dx\,dy\,=\exp \left\{ -\frac{\lambda\hbar}{m\pi}
 \int_0^t\int_{-\infty
}^\infty \left| \frac{\partial \psi _B(0,y^{\prime },t^{\prime })}{\partial x%
}\right| ^2\,\,dy^{\prime }\,dt^{\prime }\right\} . 
\]

Measurement in time gives 
\begin{eqnarray}
{\cal J}(0,y,t)&=&\frac{\lambda \hbar}{m\pi}
\left| \frac{\partial \psi (0,y,t)}{\partial x}%
\right| ^2\\
&=&\frac{\lambda \hbar}{m\pi} \left| \frac{\partial \psi _B(0,y,t)}{\partial x}\right|
^2\exp \left\{ -\frac{\lambda\hbar}{m\pi} \int_0^t\ \int_{-\infty }^\infty \left| \frac{%
\partial \psi _B(0,y^{\prime },t^{\prime })}{\partial x}\right|
^2\,\,dy^{\prime }\,dt^{\prime }\right\}  \label{Jyt}
\end{eqnarray}
This is the probability density of a collapse of the wave function occurring
at the point $y$ on the screen at time $t$. The total current 
\begin{equation}
{\cal J}(y)=\frac{\lambda\hbar}{m\pi} \int_0^\infty \left| \frac{\partial \psi (0,y,t)}{%
\partial x}\right| ^2\,dt  \label{Jy}
\end{equation}
is the probability density that the collapse of the wave function occurs at
the point $y$ on the screen (ever).

In a real experiment the measurement is neither instantaneous nor infinite
in time. That is, an integral over a finite time interval is observed rather
than (\ref{Jyt}) or (\ref{Jy}). If a packet of particles is sent out eq.(%
\ref{Jyt}) is the probability density of Feynman trajectories that propagate
instantaneously at time $t$ into the point $(0,y)$ in the screen and is
seen as the density of light intensity on the (ideal) fluorescent screen at
time $t$. The function (\ref{Jy}) represents the cumulative (in time)
probability current density of Feynman trajectories absorbed in the wall and
is seen as the density of the trace the initial packet eventually leaves on
the screen (e.g., photographic plate).

To determine the patterns (\ref{Jyt}) and (\ref{Jy}), we have to
calculate first the two-dimensional wave function with zero boundary
condition on the $y$-axis. It can be written as 
\[ \psi _B(x,y,t)=\psi _B^1(x,t)\psi ^2(y,t), \]
where, using the method of images, we find that 
\begin{eqnarray}
\psi _B^1(x,t) &=&\int_{-\infty }^0\frac{e^{-(z-x_0)^2/2\sigma _x^2}}{\sqrt{%
2\pi i}\sigma _x}\frac{e^{-im(x-z)^2/2\hbar t}}{\sqrt{2\pi i\hbar t/m}}%
\,dz-\int_0^\infty \frac{e^{-(z+x_0)^2/2\sigma _x^2}}{\sqrt{2\pi i}\sigma _x}%
\frac{e^{-im(x-z)^2/2\hbar t}}{\sqrt{2\pi i\hbar t/m}}\,dz  \nonumber \\
&=&\int_{-\infty }^0\frac{e^{-(z-x_0)^2/2\sigma _x^2}}{\sqrt{2\pi i}\sigma _x%
}\left[ \frac{e^{-im(x-z)^2/2\hbar t}}{\sqrt{2\pi i\hbar t/m}}-\frac{%
e^{-im(x+z)^2/2\hbar t}}{\sqrt{2\pi i\hbar t/m}}\right] \,dz  \label{psi1}
\end{eqnarray}
and 
\[
\psi ^2(y,t)=\int_{-\infty }^\infty \frac{e^{-z^2/2\sigma _y^2}}{\sqrt{2\pi i%
}\sigma _y}\frac{e^{-im(y-z)^2/2\hbar t}}{\sqrt{2\pi i\hbar t/m}}\,dz. 
\]
Evaluation of the integral gives 
\[
\left| \psi ^2(y,t)\right| ^2=\frac 1{2\pi \sigma _y}\left( \displaystyle{%
\frac{\hbar ^2t^2}{\sigma _y^2m^2}+\sigma _y^2}\right) ^{-1/2}\exp \left\{ {%
\displaystyle{-\frac{y^2}{\displaystyle{\frac{\hbar ^2t^2}{\sigma _y^2m^2}%
+\sigma _y^2}}}}\right\} 
\]
If $\sigma _x<<|x_0|$, the upper limit of integration in eq.(\ref{psi1}) can
be replaced by $\infty $ with a transcendentally small error. Thus, we write 
\[
\psi (x,y,t)\cong \psi ^2(y,t)\int_{-\infty }^\infty \frac{%
e^{-(z-x_0)^2/2\sigma _x^2}}{\sqrt{2\pi i}\sigma _x}\left[ \frac{%
e^{-im(x-z)^2/2\hbar t}}{\sqrt{2\pi i\hbar t/m}}-\frac{e^{-im(x+z)^2/2\hbar
t}}{\sqrt{2\pi i\hbar t/m}}\right] \,dz. 
\]

According to eq.(\ref{Jyt}), the instantaneous absorption rate at time $t$
at a point $(0,y)$ on the screen is given by 
\begin{equation}
{\cal J}(0,y,t)=\frac{4x_0^2\lambda m}{2\pi \sigma _x^2}\left( \frac{\hbar
^2t^2}{\sigma _x^2m^2}+\sigma _x^2\right) ^{-3/2}\exp \left\{ \displaystyle{-%
\frac{x_0^2}{\displaystyle{\frac{\hbar ^2t^2}{\sigma _x^2m^2}+\sigma _x^2}}}%
\right\} \left| \psi ^2(y,t)\right| ^2.  \label{absrate}
\end{equation}

To compare eq.(\ref{absrate}) with that given in \cite{Feynman}, we
reproduce the derivation of \cite{Feynman} with an initial two-dimensional
Gaussian wave packet. The result gives the wave function as 
\[
\psi _F(x,y,t)=\psi _F^1(x,t)\psi ^2(y,t) 
\]
and probability density at the screen at time $t$ as 
\begin{equation}
\left| \psi _F(0,y,t)\right| ^2=\frac 1{2\pi \sigma _x}\left( \frac{\hbar
^2t^2}{\sigma _x^2m^2}+\sigma _x^2\right) ^{-1/2}\exp \left\{ \displaystyle{-%
\frac{x_0^2}{\displaystyle{\ \frac{\hbar ^2t^2}{\sigma _x^2m^2}+\sigma _x^2}}%
}\right\} \left| \psi ^2(y,t)\right| ^2.  \label{psiF0}
\end{equation}
The instantaneous intensity of the diffraction pattern in the absence of an
absorbing screen, given in \cite{Feynman}, is defined as $\left| \psi
_F(0,y,t)\right| ^2$. Thus, the introduction of an absorbing screen,
according to these interpretations, gives the relative brightness as 
\[
\displaystyle{\frac{\displaystyle\frac{\lambda\hbar}{m\pi} \displaystyle
{\left| \displaystyle{\frac{\partial \psi (0,y,t)}{\partial x}}\right| ^2}}{%
\left| \psi _F(0,y,t)\right| ^2}}=\frac{4x_0^2\lambda }{m^2\sigma _x^6\left( %
\displaystyle{\frac{\hbar ^2t^2}{\sigma _x^2m^2}+\sigma _x^2}\right) }. 
\]
The decay in time of the quotient reflects the fact that the absorbing
screen depresses the entire wave function in time. Thus, the part of the
packet that arrives later is already attenuated by the preceding absorption,
relative to the unattenuated wave function in the absence of absorption.

Next, we propose a simple device for performing a measurement of times of
arrival of particles at an absorbing boundary in one dimension, as well as
that of the unidirectional current at the absorbing boundaries. In addition,
the proposed device demonstrates the effect of additional absorbing
boundaries on the slit experiment. The proposed measurement can discriminate
between the various modes of absorption described above.

Consider the setup of the slit experiment enclosed between two parallel
absorbing walls, symmetric with respect to the slit and perpendicular to the
planes of the screen and the slit. For example, the walls can be made of
photographic plates. Particles are given a constant initial velocity, $v_x$,
in the $x$ direction (perpendicular to the planes of the screen and slit),
within the constraints of uncertainty. The time a particle leaves the slit
is also measurable within the constraints of uncertainty.

The initial packet is Gaussian in the $x$ direction and is uniform inside
the slit (in the $y$ direction). This means that the initial velocities in
the $y$ direction have the density 
\begin{equation}
\left| \hat{\Psi}\left( k\right) \right| ^2=\left| \frac{\sin \frac \pi 2k}{%
\frac \pi 2k}\right| ^2.  \label{velocityd}
\end{equation}
The plane of the slit is $x=x_0$, the plane of the screen is $x=0$, the slit
is the interval $-\pi /2<y<\pi /2$. The absorbing planes are $y=\pm y_0$
with $y_0>\pi /2.$ In this setup the motion of the particles in the $x$
direction is independent of that in the $y$ direction. The latter is the
object of the proposed experiment.

Particles that hit the planes $y=\pm y_0$ leave traces at points $%
x_1,x_2,...,x_N$. These distances are proportional (within the constraints
of uncertainty) to the times of arrival at the absorbing walls of particles
that start out in the interval $-\pi /2<y<\pi /2$ with initial velocities
distributed as in eq.(\ref{velocityd}). The histogram obtained from these
points, on an axis normalized with the velocity $v_x,$ is that of the times
of arrival of one dimensional particles moving on the $y$ axis.

The wave function for this configuration is given by 
\[
\psi \left( x,y,t\right) =\psi ^1(x,t)\psi ^2\left( y,t\right) , 
\]
where $\left| \partial \psi ^1(0,t)/\partial x\right| ^2$ was
calculated in \cite{PAPER2} and 
\[
\psi ^2\left( y,t\right) =\sum_{n=1}^\infty \frac 2{n\pi ^{3/2}}\cos \frac{%
n\pi ^2}{4y_0}\sin \frac{n\pi }{y_0}y\exp \left\{ -\frac i\hbar \frac{n^2\pi
^2}{y_0^2}t\right\} . 
\]

According to eq.(\ref{PtautF}), the histogram of the times of arrival is
also that of the unidirectional current on each wall. According to MQM, this
pattern is given by 
\begin{eqnarray}
{\cal J}(x,\pm y_0) &=&\int_0^\infty {\cal J}(x,\pm y_0,t)\,dt  \label{Jxy0}
\\
&=&\int_0^\infty \left| \frac{\partial \psi ^1(x,t)}{\partial x}\right|
^2\left| \psi ^2(\pm y_0,t)\right| ^2\,dt.  \nonumber
\end{eqnarray}

Next, we consider the pattern observed on the wall $x=0$, $-\pi /2<y<\pi /2$%
. The histogram of the traces of the particles on the screen at $x=0$ is
given by 
\begin{equation}
{\cal J}(y)=\int_0^\infty \left| \partial \psi ^1(0,t)/\partial x\right|
^2\left| \sum_{n=1}^\infty \frac 2{n\pi ^{3/2}}\cos \frac{n\pi ^2}{4y_0}\sin 
\frac{n\pi }{y_0}y\exp \left\{ -\frac i\hbar \frac{n^2\pi ^2}{y_0^2}%
t\right\} \right| ^2\,dt.  \label{Jyx0}
\end{equation}
If the velocities in the $x$ direction are concentrated around $v_x$, the
histogram will be approximately 
\begin{eqnarray*}
{\cal J}(y) &\approx &{\cal J}(y,\bar{t})=\left| \partial \psi ^1(0,\bar{t}%
)/\partial x\right| ^2\psi ^2\left( y,\bar{t}\right) \\
&=&\left| \partial \psi ^1(0,\bar{t})/\partial x\right| ^2\left|
\sum_{n=1}^\infty \frac 2{n\pi ^{3/2}}\cos \frac{n\pi ^2}{4y_0}\sin \frac{%
n\pi }{y_0}y\exp \left\{ -\frac i\hbar \frac{n^2\pi ^2}{y_0^2}\bar{t}%
\right\} \right| ^2.
\end{eqnarray*}
This is not the same as the expression obtained in eq.(\ref{Jy}). The
difference is due to the effect of the lateral absorbing boundaries. Thus,
according to MQM, absorbing boundaries cannot be ignored, as done in QM.\\

\noindent
{\large{\bf 7. Recovering the free particle probability density function from
measurements of unidirectional currents}}\\
\renewcommand{\theequation}{7.\arabic{equation}}\setcounter{equation}{0}

The wave function of a freely propagating initial Gaussian packet (Gaussian
slit) can be evaluated from measurements by a detector in the form of an
absorbing screen. It suffices to rotate the screen to obtain two
measurements that reproduce the entire wave function of the free packet as
follows.

It was shown above that $\left| \psi _F(0,y,t)\right| ^2$ can be recovered,
up to a multiplicative function of time, from measurements of the
unidirectional current in the Gaussian slit experiment, 
\[
\left| \psi _F(0,y,t)\right| ^2=\frac{{\cal J}(0,y,t)}{\phi (t)}, 
\]
where eq.(\ref{psiF0}) gives 
\[
\phi (t)=\frac{4x_0^2\lambda m}{2\pi \sigma _x^2}\left( \frac{\hbar ^2t^2}{%
\sigma _x^2m^2}+\sigma _x^2\right) ^{-3/2}\exp \left\{ \displaystyle{-\frac{%
x_0^2}{\displaystyle{\frac{\hbar ^2t^2}{\sigma _x^2m^2}+\sigma _x^2}}}%
\right\} . 
\]
Now, we rotate the screen plane about the point $\left( 0,0\right) $ on the
screen by an angle $\theta $ and introduce a new coordinate system about the
same origin, 
\begin{equation}
x^{\prime }=\alpha x+\beta y,\quad y^{\prime }=\gamma x+\delta y,
\label{x'y'}
\end{equation}
where $\alpha =\cos \theta ,\,\beta =\sin \theta ,\,\gamma =-\sin \theta
,\,\delta =\cos \theta $. We write 
\[
\left| \psi _F(x,y,t)\right| ^2=\left| \tilde{\psi}_F(x^{\prime },y^{\prime
},t)\right| ^2 
\]
so that 
\begin{equation}
\psi _F^1(x,t)\psi ^2(y,t)=\tilde{\psi}_F^1(x^{\prime },t)\tilde{\psi}%
^2(y^{\prime },t).  \label{psiFpsiF'}
\end{equation}
According to our theory, 
\[\left| \tilde{\psi}^2(y',t)\right| ^2=\frac{\tilde{\cal J}
(a',y',t)}{\tilde{\phi}(t)}, \]
where $a^{\prime }$ is the distance from the center of the initial packet
to the rotated screen and 
\[
\tilde{\phi}(t)=\frac{4a^{\prime 2}\lambda m}{2\pi \sigma _x^2}\left( \frac{%
\hbar ^2t^2}{\sigma _x^2m^2}+\sigma _x^2\right) ^{-3/2}\exp \left\{ %
\displaystyle{-\frac{a^{\prime 2}}{\displaystyle{\frac{\hbar ^2t^2}{\sigma
_x^2m^2}+\sigma _x^2}}}\right\} . 
\]
The current $\tilde{\cal J}(a',y',t)$ is given by 
\[
\tilde{\cal J}(a',y',t)=\frac{4a^{\prime 2}\lambda m}{%
2\pi \sigma _x^2}\left( \frac{\hbar ^2t^2}{\sigma _x^2m^2}+\sigma
_x^2\right) ^{-3/2}\exp \left\{ \displaystyle{-\frac{a^{\prime 2}}{%
\displaystyle{\frac{\hbar ^2t^2}{\sigma _x^2m^2}+\sigma _x^2}}}\right\}
\left| \tilde{\psi}^2\left( y^{\prime },t\right) \right| ^2. 
\]
It should be borne in mind that ${\cal J}(0,y,t)$ and $\tilde{\cal J}%
(a^{\prime },y^{\prime },t)$ are measurable on the screen.

Now, eqs.(\ref{psiFpsiF'}) and (\ref{x'y'}) give 
\[
\left| \tilde{\psi}_F^1(a^{\prime },t)\tilde{\psi}^2(y^{\prime },t)\right|
^2=\left| \psi _F^1\left( \frac{a^{\prime }-\beta y}\alpha ,t\right) \psi
^2(y,t)\right| ^2. 
\]
It follows that 
\[
\left| \psi _F^1\left( \frac{a^{\prime }-\beta y}\alpha ,t\right) \right| ^2=%
\frac{\tilde{\cal J}\left( a^{\prime },\gamma \frac{a^{\prime }-\beta y}%
\alpha +\delta y,t\right) \phi (t)\left| \tilde{\psi}_F^1(a^{\prime
},t\right| ^2}{{\cal J}(0,y,t)\tilde{\phi}(t)}, 
\]
hence 
\begin{eqnarray}
\left| \psi _F\left( x,y,t\right) \right| ^2 &=&\left| \psi _F^1(x,t)\psi
^2(y,t)\right| ^2  \label{psiF2} \\
&=&\frac{\tilde{\cal J}\left( a^{\prime },\gamma x+\delta \frac{a^{\prime
}-\alpha x}\beta ,t\right) {\cal J}(0,y,t)\left| \tilde{\psi}_F^1(a^{\prime
},t\right| ^2}{{\cal J}\left( 0,\frac{a^{\prime }-\alpha x}\beta ,t\right) 
\tilde{\phi}(t)}.  \nonumber
\end{eqnarray}
Thus, up to a time dependent normalization factor, the wave function in the
absence of the screen can be found from measurable currents on the two
screens. Since 
\[
\int \int \left| \psi _F\left( x,y,t\right) \right| ^2\,dx\,dy=1, 
\]
we can rewrite eq.(\ref{psiF2}) as 
\[
\left| \psi _F\left( x,y,t\right) \right| ^2={\cal N}^{-1}\left( t\right) 
\frac{\tilde{\cal J}\left( a^{\prime },\gamma x+\delta \frac{a^{\prime
}-\alpha x}\beta ,t\right) {\cal J}(0,y,t)}{{\cal J}\left( 0,\frac{a^{\prime
}-\alpha x}\beta ,t\right) }, 
\]
where 
\[
{\cal N}^{-1}\left( t\right) =\int \int \frac{\tilde{\cal J}\left(
a^{\prime },\gamma x+\delta \frac{a^{\prime }-\alpha x}\beta ,t\right) {\cal %
J}(0,y,t)}{{\cal J}\left( 0,\frac{a^{\prime }-\alpha x}\beta ,t\right) }%
\,dx\,dy. 
\]

\noindent
{\large{\bf 8. Summary}} \\
\renewcommand{\theequation}{8.\arabic{equation}}\setcounter{equation}{0}

A measuring device of the coordinate of a particle is modeled a an
absorbing wall that cuts off the wave function. This is a more
physically realistic model than von Neumann's collapse machine or other
quantum collapse devices. The proposed detector modifies the wave
function of the particle in an irreversible manner. It replaces the QM
collapse of the wave function with the usual collapse of the
probability distribution after observation. The proposed MQM is based
on postulates that assume well defined trajectories of quantum
particles. The calculation of the wave function of a particle in the
presence of such detector is based on the Feynman integral and a
discounting procedure of the probability of absorbed trajectories. The
resulting MQM is Hamiltonian prior to absorption but the absorption
process is not. The fundamental notion in MQM is the absorption
probability current on the absorbing wall. This is used to define the
probability distribution of the arrival time and of the arrival
point.\\

\end{document}